*1.1. CENTERIS - International Conference on ENTERprise Information Systems*

# Towards a Collaborative Approach to Decision Making Based on Ontology and Multi-Agent System

# Application to crisis management

Ahmed Maalel [a,b,*] and Henda Ben Ghézala[a]

[a] *University of Manouba, National School of Computer Sciences, RIADI Laboratory, 2010, Manouba, Tunisia*
[b] *University of Sousse, Higher Institute of Applied Science and Technology, 4003, Sousse, Tunisia*

*{ahmed.maalel, henda.benghezala}@ensi.rnu.tn*

**Abstract**

The coordination and cooperation of all the stakeholders involved is a decisive point for the control and the resolution of problems. In the insecurity events, the resolution should refer to a plan that defines a general framework of the procedures to be undertaken and the instructions to be complied with; also, a more precise process must be defined by the actors to deal with the case represented by the particular problem of the current situation. Indeed, this process has to cope with a dynamic, unstable and unpredictable environment, due to the heterogeneity and multiplicity of stakeholders, and finally due to their possible geographical distribution. In this article, we will present the first steps of validation of a collaborative decision-making approach in the context of crisis situations such as road accidents. This approach is based on ontologies and multi-agent systems.

## 1. Introduction

Generally, in the crisis management, various stakeholders (Experts, Police, technical investigators, hospitals, citizens, etc.) are called upon in order to reduce criticality and the impact of the situation produced. The coordination of these actors is a determining point for the control and the resolution of problems. This coordination is ensured by a steering group, which is composed of an institutional manager and representatives of the involved stakeholders. To solve an insecurity event, a specific process must be defined by the actors to deal with the particular case of the current situation. Moreover, this process must cope with a dynamic, unstable and unpredictable environment, due to the consequences and evolution of the event of insecurity over time, but also to the heterogeneity and multiplicity of stakeholders. This incomplete and evolving process needs to be revised, adapted or refined to consider the human, institutional and collective dimension and to integrate interaction techniques to facilitate decision-making. In conclusion, it would be appropriate to propose an interactive and collaborative approach to crisis management that makes it possible to apprehend the shortcomings already mentioned. Through a practical and intelligent solution allowing the most informed and adapted decision-making. In this article, we will present the first steps of validation of a collaborative decision-making approach in the context of crisis situations such as road accidents. This approach is based on ontologies and multi-agent systems. The organization of this paper is as follows. The second section presents a review of related works. The next section describes our suggested approach, through the conceptual model, the domain ontology, and the agent's architecture. Finally, the last section presents an example of integration of ontology into a Mutli-Agent system.

## 2. Related Works

Several studies have been proposed in the area of crisis management. We include below a table which compares the different approaches, according to the criteria that we have set.

Table 1. Summary of related work

| Ref | Name | Domain | Knowledge Representation | | Knowledge exploitation | | Integration |
| --- | --- | --- | --- | --- | --- | --- | --- |
| | | | Classical formalism (DB, Text, etc.) | Semantics methods | MAS | Others | |
| [1] | IsyCri | All crisis | | × | | | |
| [2] | | Terrorism | | × | | | |
| [3] | | Shnow Storm | | × | | | |
| [4] | | All crisis | | × | | | |
| [5] | Adast | Railroad Accident | | × | | CBR | |
| [6] | SimGenis | Accident | × | | × | | |
| [7] | WIPER | Generic | × | | × (mobile) | | |
| [8] | ICrisis | Generic | × | | | | |
| [9] | | Generic | × | | × | CBR | |
| [10] | SimlCrise | Maritime Accident | × | | × | | |
| [11] | | Natural Disaster | × | | × | | |
| [12] | | Snow Storm | × | | × (BDI) | | |
| [13] | | Air pollution | × | | × | ANN | |
| [14] | | Artisan creation | | × | × | | SPARQL |
| [15] | | Urban logistics | | × | × | | SPARQL |
| [16] | | Search on the Web | | × | × | | SPARQL |
| [17] | | Firefighting | | × | × | | |
| [18] | SWIMS | Emergency traffic | | × | × (BDI) | Web Services | OBG |
| [19] | SIMFOR | All crisis | | × | × (BDI) | Serious Games | |
| [20] | | Aircraft accident | | × | × (BDI) | | |

The main points mentioned in the table are presented below:

- Application domain: There exist two types of domain application: generic or specific.
- *Knowledge representation:* Traditional (Data base, text, etc.) or Semantic methods (ontologies, etc.)
- *Knowledge exploitation:* traditional or advanced methods (CBR, MAS, etc.)
- *Integration:* tool of integration between Ontology and MAS: SPARQL or OBG (Ontology Bean Generator).

Despite the important efforts made in order to propose approaches to the decision based on the MAS and ontologies, and although these works make significant contributions, some limitations can be detected. We note that all ontologies are almost domain ontologies, it turns out that their deployment in other areas rather than the area for which it has been designed is difficult or impossible. In addition, a key part in our research is the crisis management phases. We also reveal that most ontologies cover only one phase, generally the operational phase. In fact, few ontologies support coordination between the actors involved and represent the roles and interactions between them. Furthermore, almost all ontologies lack mechanisms and crisis prevention methods. They need deeper consideration and improvements. In most of the works presented, the practices are classical; we note a lack and insufficiency of works using semantic practices. In spite of all efforts to propose approaches and techniques to manage crises, most suffer from the lack of formalism to express or model the knowledge about crisis scenarios. At the knowledge exploitation level, most approaches using MAS do not really integrate all stakeholders. In fact, they focus only on special category of actors. Thus, roles are usually fixed and therefore the actors cannot support changes of environment and events. Indeed, there are problems of coordination, problems of communication between the actors at the time of the crisis and the lack of respect of the responsibilities between them.

## 3. Proposed Approach

The proposed approach is structured around a specific sequence of steps:

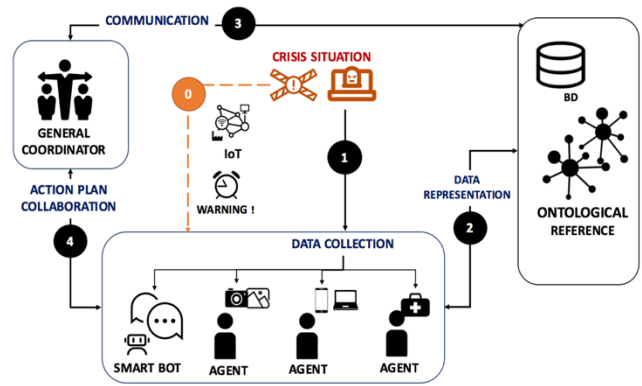

Fig. 1. Our envisaged process for collaborative and context-aware mobile decision support approach for real-time emergency management.

- Step 0: Crisis detection and analysis phase, occurs when a warning signs of crisis are recognized and sending it to an agent for analysis and verification of the information source;
- Step 1: Crisis selection phase, when a crisis is verified, a crisis core ontology is generated in order to select convenient types and match it with specific ontologies for a deep understanding of the crisis severity;
- Step 2 Situation awareness building phase, consist on the generation of context ontology for the purpose of adding additional knowledge on climate, geography, citizens, damage level. Matching context with crisis ontologies assist various actors to get a clearer picture of the crisis scene as well as taking all necessary precautions;

- Step 3: Assembling and solution generation phase is a blend of ontologies (tasks, methods, process, resource, users, plans and strategies…) and collaboration patterns assembled by a context ontology to produce a cooperation and communion medium to manage crisis;
- Step 4: Decision making phase is a reasoning process for optimal decision support to all users (fixed or mobile) pursued by a monitoring phase to go after the efficiency of the generated solution.

## 3.1. The knowledge representation

Now, we are immigrating from conceptual model to a model implemented in one of ontology languages such as OWL, RDF etc. To implement it, we have used the ontology editor Protege. The ontology is used to explain the different concepts of crisis management (terrorist attack) in a generic and comprehensive way. This ontology represents different kinds of knowledge: crisis's phases (before, during and after), missions and roles, the environment entities (actors, situation's type, context, consequences ...) and the different interaction between elements (interactions). The below figure 2 shows the class hierarchy of the ontology based on our domain model

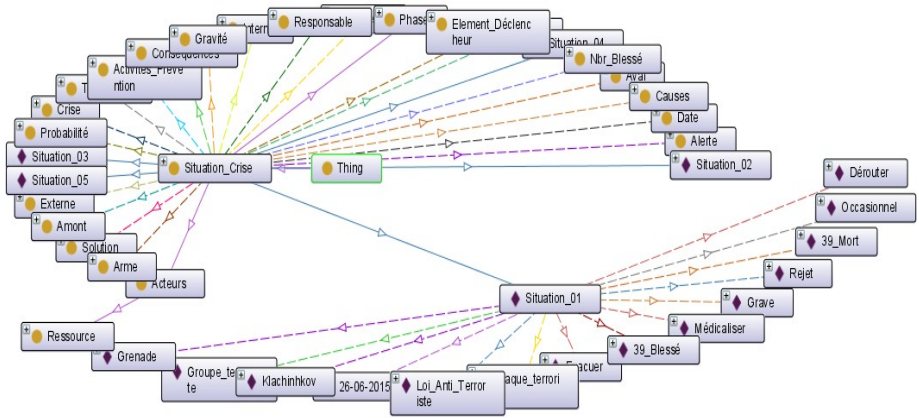

Fig.2. Class hierarchy of the domain ontology

## 3.2. Feasibility model and Validation

Very schematically, to better manage a crisis situation and therefore make the most appropriate decision, it was necessary to ensure a very good collection of information on the crisis situation that occurs, and this, in real time. The coordinating agent (decision maker) uses several operator agents. Each agent, has a specific task. Operator agents send the information in real time to the coordinator, who will process this data and choose an appropriate decision for the situation and then make a recommendation to operators who are observers. So, the role of the coordinating agent who is the decision maker is to ensure the sharing of tasks and results. He has several tasks to ask others to do a game, then he has to integrate the result. The crisis situation will be distributed among several agents, and these agents are cooperative agents who communicate with the coordinating agent. Two-way communication is provided by the FIPA-ACL protocol. We have set up an ontology and a model based on MAS to manage road accidents. We integrated domain ontology with Jena Framework and SPARQL. To validate our contribution, we are dealing with a concrete case of a road accident. Figure 3 illustrates the actors involved in this case of experimentation. The model makes use of three agents: decision maker, observer agent 1, observer agent 2 and camera agent.

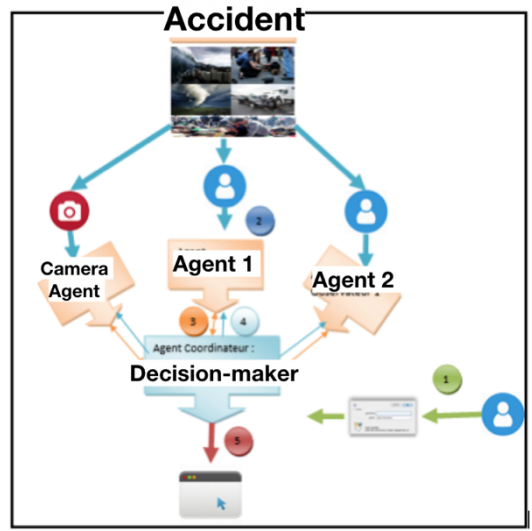

Fig. 3 Functional architecture of the developed prototype

The sequence is as follows:

1) Decision maker authentication and agent deployment.
2) The observer agent 1 and observer agents describe the crisis situation and the camera agent takes pictures in real time.
3) The decision-maker requests the information collected by the agents.
4) The decision-maker processes this data and displays it in its interface.
5) Finally, the decision-maker makes a recommendation to the observers.

So that the decision-maker can make a reliable decision for the current crisis situation, he needs a stream of photos in real time. For this we will use a smartphone connected to the internet. The figure below shows a screenshot of the Droid Cam application in the smartphone:

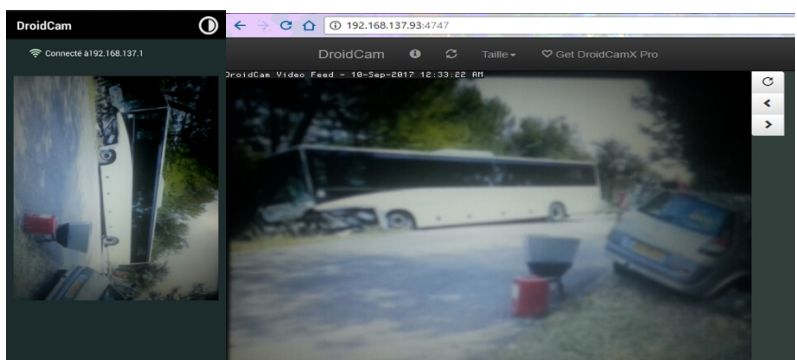

Fig. 4. Screen Prints of the DroidCam and Web Application

The decision-making agent collects all the data received from the agents, processes them and displays them in his interface in order to manage the situation and make a recommendation to the observer agent 2. The figure below is the Sniffer Agent Interface which is a tool in the JADE platform. This interface presents the diagram of interaction between all the deployed agents in our prototype.

## 4. Conclusion

Generally, crisis systems include heterogeneous, distributed and dynamic actors. They have to interact frequently and cooperate at a high level to deal with the crisis. We have presented in this paper a feasibility model of decision making to improve coordination between involved entities. Our approach is to combine multi-agent problem solving and better decision-making. We expect in terms of prospects:

- The possibility of using other connected objects like RFID sensor, Drone, robot, etc. to ensure a good collection of information in real time.
- To link the application to other emergency response agents such as Emergency Medical Services with Emergency Medical Assistance Services (EMS) supported by Emergency and Emergency Medical Services. Resuscitation (SMUR), health services officer, army officer and national police officer etc.